# Effects of Neutrino Oscillation on the Supernova Neutrino Spectrum

Keitaro Takahashi*, Mariko Watanabe* and Katsuhiko Sato*,**

*Department of Physics, School of Science, the University of Tokyo,
Hongo 7-3-1, Bunkyo-ku, Tokyo 113-0033, Japan
**Research Center for the Early Universe, School of Science, the University of Tokyo,
Hongo 7-3-1, Bunkyo-Ku, Tokyo 113-0033, Japan

and

Tomonori Totani

Theory Division, National Astronomical Observatory,
Mitaka, Tokyo 181-8588, Japan

The effects of three-flavor neutrino oscillation on the supernova neutrino spectrum are studied. We calculate the expected event rate and energy spectra, and their time evolution at the Superkamiokande (SK) and the Sudbury Neutrino Observatory (SNO), by using a realistic neutrino burst model based on numerical simulations of supernova explosions. We also employ a realistic density profile based on a presupernova model for the calculation of neutrino conversion probability in supernova envelopes. These realistic models and numerical calculations allow us to quantitatively estimate the effects of neutrino oscillation in a more realistic way than previous studies. We then found that the degeneracy of the solutions of the solar neutrino problem can be broken by the combination of the SK and SNO detections of a future Galactic supernova.

## I. INTRODUCTION

Neutrinos are the most mysterious particle in the standard model of particle physics, and the only elementary particles showing evidence for new physics beyond the standard model. While the evidence for the existence of neutrino oscillations from solar [1,2] and atmospheric neutrino [3] data is rather convincing now, the values of the mass squared differences and mixing angles are not firmly established. For the observed $\nu_e$ suppression of solar neutrinos, for example, four solutions are still possible: large mixing angle (LMA), small mixing angle (SMA), low $\Delta m^2$ (LOW), and vacuum oscillation (VO). For $\theta_{13}$, the mixing angle between mass eigenstate $\nu_1, \nu_3$, only upper bound is known from reactor experiment [4] and combined three generation analysis [5]. Also the nature of neutrino mass hierarchy (normal or inverted) is still a matter of controversy.

There is another neutrino source: supernovae. This is a completely different system from solar, atmospheric, accelerator, and reactor neutrinos in regard to neutrino energy and flavor of produced neutrinos, propagation length and so forth. Then neutrino emission from a supernova is expected to give valuable information that can not be obtained from neutrinos from other sources.

Neutrino astrophysics entered a new phase when neutrinos from SN1987A in the Large Magellanic Cloud were detected by the Kamiokande [6] and IMB [7]. These pioneering observations contributed significantly to our knowledge of the fundamental properties of neutrinos [8–12] as well as our understanding of the mechanisms involved in a collapse-driven supernova. However, event numbers at the Kamiokande and IMB, 11 and 8 events respectively, are too small to set statistically robust constraints.

The next Galactic supernova will be even more valuable because of the abundance of neutrino events produced by a closer source and detected by new neutrino detectors which are now available. For example, Kamiokande has been upgraded into SuperKamiokande(SK), whose volume is about 15 times larger than the old one. Sudbury Neutrino Observatory (SNO), which is unique in its use of heavy water, has already been in operation [13].

There have been some studies on future supernova neutrino detection taking neutrino oscillation into account. Dighe and Smirnov [14] estimated qualitatively the effects of neutrino oscillation in a collapse-driven supernova on the neutronization peak, the distortion of energy spectra, and the Earth matter effects. They concluded that it is possible to identify the solar neutrino solution and to probe the mixing angle $\theta_{13}$. Dutta et al. [15] showed numerically that the events involving oxygen targets increase dramatically when there is neutrino mixing.

In this paper, we calculate numerically the effects of three flavor oscillation on the supernova neutrino spectra, taking into account the constraints on the neutrino mixing and masses imposed by solutions consistent with the solar and atmospheric neutrino problems. For the original neutrino flux and density profile of the supernova, which were not considered accurately in previous studies, we use ones which are based on the realistic numerical presupernova and supernova models. We then calculate number of events expected to be detected at SuperKamiokande and SNO. Finally we propose a method to discriminate quantitatively the solutions of the solar neutrino problem. We do not include the Earth matter effects in our calculation, and this point will be discussed when we conclude the paper.

This paper is organized as follows. In section II we describe the features of supernova neutrinos. In section III we calculate dynamics of neutrino conversion on their way out to the surface of the star. Then, in section IV, we obtain neutrino energy spectra at detectors and time evolution of neutrino number luminosity. Features of results of section IV are discussed in section V and a measure of neutrino oscillation is proposed. Finally in section VI, we summarize our results.

## II. SUPERNOVA NEUTRINOS

Supernova neutrino emission process can be divided into two distinct phases [16]; the neutronization burst and nearly thermal neutrino emission. Almost all of the binding energy,

$$E_b = 1.5 \sim 4.5 \times 10^{53} \text{erg}, \qquad (2.1)$$

is radiated away as neutrinos, while a small fraction of which ($\sim 2 \times 10^{51}$ erg) are emitted during the first phase. While only $\nu_e$ is emitted during the first phase, neutrinos and antineutrinos of all types are emitted during the second phase with rougly the same luminosity. The average energies are different between flavors:

$$\langle E_{\nu_e} \rangle \simeq 13 \text{MeV} \qquad (2.2)$$
$$\langle E_{\bar{\nu}_e} \rangle \simeq 16 \text{MeV} \qquad (2.3)$$
$$\langle E_{\nu_x} \rangle \simeq 23 \text{MeV}, \qquad (2.4)$$

where $\nu_x$ means either of $\nu_\mu$, $\nu_\tau$, and their antineutrinos. In this paper we use a realistic model of a collapse-driven supernova by the Lawrence Livermore group [17] to calculate the neutrino luminosity and energy spectrum. Time-integrated energy spectra and time evolution of neutrino flux are are shown in Fig. 1 and Fig. 2. (See Totani et al. [18] for detail.)

These neutrinos, which are produced in the high dense region of the iron core, interact with matter before emerging from the supernova. The presence of non-zero masses and mixing in vacuum among various neutrino flavors results in strong matter dependent effects, including conversion from one flavor to another. In supernova, the conversions occur mainly in the resonance layers. The resonance matter density can be written as

$$\rho_{res} \sim 1.4 \times 10^6 (\frac{\Delta m^2}{1\text{eV}^2})(\frac{10\text{MeV}}{E})(\frac{0.5}{Y_e}) \cos\theta \text{ g/cc}, \quad (2.5)$$

where$\Delta m^2$ is the mass squared difference, $E$ is the neutrino energy, and $Y_e$ is the mean number of electrons per baryon. In normal mass hierarchy scheme ($m_3 > m_2 > m_1$), the system has two resonances in neutrino sector: one at higher density(H-resonance) and the other at lower density(L-resonance). On the other hand, antineutrino sector has no resonance.

The dynamics of conversions in each resonance is determined by the adiabaticity parameter $\gamma$,

$$\gamma \equiv \frac{\Delta m^2}{2E} \frac{\sin^2 2\theta}{\cos 2\theta} \frac{n_e}{dn_e/dr}, \qquad (2.6)$$

where $\theta$ is mixing angle, and $n_e$ is the electron number density. The flip probability $P_f$, the probability that a neutrino in one matter eigenstate jumps to the other matter eigenstate, is,

$$P_f = \exp(-\frac{\pi}{2}\gamma), \qquad (2.7)$$

as given by the Landau-Zener formula [19]. Adiabatic resonance corresponds to $\gamma \gg 1$. Note that adiabaticity of resonance depends on the mixing angle and the squared mass difference, that is [14],

$$\text{H resonance} \longrightarrow \theta_{13}, \Delta m^2_{13}, \qquad (2.8)$$
$$\text{L resonance} \longrightarrow \theta_{12}, \Delta m^2_{12}. \qquad (2.9)$$

The neutrino spectra observed at the detectors can be dramatically different from the original spectra according to the adiabaticity of these two resonances. If neutrino oscillation occurs, for example, between $\nu_e$ and $\nu_x$, observed energy spectrum will be a mixture of original $\nu_e$ and $\nu_x$ spectra and the average energy of $\nu_e$ will be higher than the original $\nu_e$ average energy.

We use the massive star density profile calculated numerically by Woosley and Weaver [20] to calculate the time evolution of neutrino wave functions. The progenitor mass was set to be $15M_\odot$, and the metallicity was set to be the same as that of Sun. The density profile is shown in Fig. 3.

## III. NUMERICAL CALCULATION OF CONVERSION PROBABILITIES

### A. Time Evolution of Conversion Probabilities in the Framework of Three Flavor Neutrinos

In the framework of three-flavor neutrino oscillation, the time evolution equation of the neutrino wave functions can be written as follows:

$$i\frac{d}{dt}\begin{pmatrix}\nu_e \\ \nu_\mu \\ \nu_\tau\end{pmatrix} = H(t)\begin{pmatrix}\nu_e \\ \nu_\mu \\ \nu_\tau\end{pmatrix} \qquad (3.1)$$

$$H(t) \equiv U \begin{pmatrix} 0 & 0 & 0 \\ 0 & \Delta m^2_{21}/2E & 0 \\ 0 & 0 & \Delta m^2_{31}/2E \end{pmatrix} U^{-1}$$

$$+ \begin{pmatrix} A(t) & 0 & 0 \\ 0 & 0 & 0 \\ 0 & 0 & 0 \end{pmatrix}, \quad (3.2)$$

where $A(t) = \sqrt{2} G_F n_e(t)$, $G_F$ is Fermi constant, $n_e(t)$ is the electron number density, $\Delta m_{ij}^2$ is the mass squared differences, and $E$ is the neutrino energy. In case of antineutrino, the sign of $A(t)$ changes. Here U is a unitary $3 \times 3$ mixing matrix in vacuum:

$$U = \begin{pmatrix} c_{12}c_{13} & s_{12}c_{13} & s_{13} \\ -s_{12}c_{23} - c_{12}s_{23}s_{13} & c_{12}c_{23} - s_{12}s_{23}s_{13} & s_{23}c_{13} \\ s_{12}s_{23} - c_{12}c_{23}s_{13} & -c_{12}s_{23} - s_{12}c_{23}s_{13} & c_{23}c_{13} \end{pmatrix},$$
(3.3)

where $s_{ij} = \sin \theta_{ij}, c_{ij} = \cos \theta_{ij}$ for $i, j = 1, 2, 3 (i < j)$. We have here put the CP phase equal to zero in the CKM matrix.

In $H(t)$, the first term is the origin of vacuum oscillation, and the second term A(t), which is the only time-dependent term in H(t), is the origin of MSW effect.

By solving numerically the above differential equations from the center of supernova to the outside of supernova, we obtain conversion probabilities $P(\nu_{\alpha \to \beta})$, i.e., probabilities that a neutrino of flavor $\alpha$ produced at the center of supernova is observed as a neutrino of flavor $\beta$.

We assume normal mass hierarchy and use the sets of mixing parameters shown in table I. Here $\theta_{12}$ and $\Delta m_{12}^2$ correspond to the solutions of solar neutrino problem and $\theta_{23}$ and $\Delta m_{13}^2$ correspond to the solution of atmospheric neutrino. The value of $\theta_{13}$ is taken to be consistent with current upper bound from reactor experiment [4]. These models are named after their values of mixing angle: LMA-L means that $\theta_{12}$ is set to be LMA of solar neutrino problem and $\theta_{13}$ is large.

Fig.4 - 7 show the time evolution of coversion probability. $P(e \to e)$ and $P(e \to x)$ means the probability that $\nu_e$ produced at the center of supernova become $\nu_e$, $\nu_x(\nu_\mu, \nu_\tau)$, respectively. Since we set $\sin^2 2\theta_{23} = 1$, probabilities to become $\nu_\mu$ and $\nu_\tau$ are the same. Four lines correspond to a neutrino of energy, 5MeV, 10MeV, 40MeV, 70MeV, respectively.

As can be seen, for example, H resonance occurs adiabatically at $r = 0.02 \sim 0.05 R_\odot$ (O+Ne+Mg or O+C layer), since $\theta_{13}$ is large in model LMA-L. Final conversion probabilities are independent of neutrino energy because of the adiabaticity of resonance. On the other hand, in the lower of Fig 6 while H resonance is adiabatic, L resonance (He layer) is nonadiabatic because $\theta_{12}$ is small. Consequently, final conversion probabilities depend on energy. More detailed study on dynamics of conversion probability in supernovae is done by Watanabe [21].

We also calculate conversion probabilities with parameter sets which correspond to the LOW and VO solutions of solar neutrino problem:

$$(\text{LOW}) \Delta m_\odot^2 \approx (0.5 \sim 2) \times 10^{-7} \text{eV}^2 \quad (3.4)$$

$$\sin^2 2\theta_\odot \approx 0.9 \sim 1.0 \quad (3.5)$$
$$(\text{VO}) \Delta m_\odot^2 \approx (0.6 \sim 6) \times 10^{-10} \text{eV}^2 \quad (3.6)$$
$$\sin^2 2\theta_\odot \approx 0.8 \sim 1.0 \quad (3.7)$$

In these cases, whose parameters have nearly the same values as LMA except for $\Delta m_{12}^2$, the final conversion probabilities in vacuum take nearly the same values as in the case of LMA. The difference in $\Delta m_{12}^2$ is reflected in the radius at which L resonance occurs. Larger $\Delta m_{12}^2$ results in lager radius.

## IV. EXPECTED EVENT RATES IN SUPERKAMIOKANDE AND SNO

In this section, expected event rates at SuperKamiokande and SNO are studied assuming a future galactic supernova at a distance d = 10kpc. To obtain the event rates, we use the original neutrino flux mentioned in section II and the probabilities of the flavor conversion calculated in the previous section. We performed smoothing of the electron/positron energy spectra with a dispersion of 1 MeV, taking into account rough energy resolutions of detectors. In fact, the energy resolutions of detectors depend on energy itself. But the rough estimation of energy resolution is sufficient, since our purpose is to see overall shapes of the spectra. We also assume that the time delay of neutrinos due to non-zero mass is negligible compared with the time scale considered here (> msec). This assumption is secure unless the neutrino mass hierarchy is degenerate at around $\gtrsim$ 3 MeV [22].

Since the original neutrino spectra and the conversion probabilities are the same for $\nu_\mu$ and $\nu_\tau$, the event rates are also the same for $\nu_\mu$ and $\nu_\tau$.

### A. Event Rates at SuperKamiokande

SuperKamiokande is a water Cherenkov detector with 32,000 ton pure water based at Kamioka in Japan. The relevant interactions of neutrinos with water are as follows:

$$\bar{\nu}_e + p \to n + e^+ \quad (\text{CC}) \quad (4.1)$$
$$\nu_e + e^- \to \nu_e + e^- \quad (\text{CC and NC}) \quad (4.2)$$
$$\bar{\nu}_e + e^- \to \bar{\nu}_e + e^- \quad (\text{CC and NC}) \quad (4.3)$$
$$\nu_x + e^- \to \nu_x + e^- \quad (\text{NC}) \quad (4.4)$$
$$\nu_e + O \to F + e^- \quad (\text{CC}) \quad (4.5)$$
$$\bar{\nu}_e + O \to N + e^+ \quad (\text{CC}) \quad (4.6)$$

where CC and NC stand for charged current and neutral current interactions, respectively. The lower limit of detection is $\sim$ 5MeV, and the energy resolution is $\sim$ 15% for an electron with energy 10MeV. For the cross sections of these interactions, we refer to [23]. The appropriate detection efficiency curve is also taken into account [24].

The efficiency is 100% above 5MeV and 50% at 4.5MeV. In these interactions, the $\bar{\nu}_e p$ CC interaction [Eq.(4.1)] has the largest contribution to the detected events at SK. Hence the energy spectrum detected at SK (including all the reactions) is almost the same as the spectrum derived from the interaction Eq.(4.1) only.

Fig. 8 and Fig. 9 show energy spectrum and time evolution of number luminosity of positrons and electrons expected to be detected at SuperKamiokande, respectively. Fig. 10 is a zoom-up of Fig. 9 near the neutronization burst.

Total event numbers for all the models are shown in Table II. In this table, the numbers of events for each interaction and contribution from neutronization burst phase are also shown. Here the neutronization phase means the period from 41msec to 48msec after the bounce.

### B. Event Rates in SNO

Sudbury Neutrino Observatory(SNO) is a water Čherenkov detector based at Sudbury, Ontario. SNO is unique in its use of 1000 tons of heavy water, by which both the charged-current and neutral-current interactions can be detected. The interactions of neutrinos with heavy water are as follows,

$$\nu_e + d \to p + p + e^- \quad \text{(CC)} \quad (4.7)$$
$$\bar{\nu}_e + d \to n + n + e^+ \quad \text{(CC)} \quad (4.8)$$
$$\nu_x + d \to n + p + \nu_x \quad \text{(NC)} \quad (4.9)$$
$$\bar{\nu}_x + d \to n + p + \bar{\nu}_x \quad \text{(NC)} \quad (4.10)$$

The two interactions written in Eqs.(4.7) and (4.8) are detected when electrons emit Čherenkov light. These reactions produce electrons and positrons whose energies sensitive to the neutrino energy, and hence the energy spectra of electrons and positrons give us the information on the original neutrino flux. In this work, we mainly take into account these two charged current interactions. For the cross sections, we refer to [25]. The efficiency of detection is set to be one, because we have no information about it.

Two neutral current interactions, which produce neutrons, are detected by observing the photons emitted at the neutron absorption. Photons give energy to electrons, then the Čherenkov light from the electrons is detected. Moreover, there is a possibility to distinguish the two CC interactions by detecting neutrons because the detection of the neutron and the positron at the same time indicates the interaction in Eq.(4.8).

Fig.11 and Fig.12 show energy spectrum and time evolution of number luminosity of positrons and electrons, produced by the two CC interactions expected at SNO, respectively. Fig.13 is a zoom-up of Fig.12 near the neutronization burst.

Total event numbers for all the models are shown in the following Table III. In this table, the numbers of events of each interaction and contribution from neutronization burst phase are also shown. Here the neutronization phase means the period from 41msec to 48msec after the bounce.

The SNO detector has also 7,000 tons of light water which can be used to detect neutrinos. This can be considered to be a miniature of SuperKamiokande (32,000 tons of light water). Then the number of events detected by light water at SNO is 7/32 of that at SuperKamiokande.

## V. DISCUSSION

### A. Features of Energy Spectra and Neutronization

As can be seen in Fig.8 and 11, when there is neutrino oscillation, neutrino spectra are harder than those in absence of neutrino oscillation. This is because average energies of $\nu_e$ and $\bar{\nu}_e$ are smaller than those of $\nu_x$ and neutrino oscillation produces high energy $\nu_e$ and $\bar{\nu}_e$ which was originally $\nu_x$. This feature can be used as a criterion of magnitude of neutrino oscillation, which will be discussed in the next section.

It is worth noting that number of events during neutronization burst phase is highly suppressed in model LMA-L and SMA-L. This is because, due to large value of $\theta_{13}$ in these two models, H resonance occurs adiabatically and $\nu_e$ produced at the center of supernova is detected as $\nu_x$ which has small cross section. But the number of events during neutronization burst will be too small to extract statistically significant information (see Table II and III).

It is possible that He and H layers of progenitor star are missing when supernova burst occurs, and density decreases abruptly outside the O+C layer. Then L resonance would occur nonadiabatically to some extent even in case of LMA, and differences between LMA and SMA would become smaller.

### B. Comparison of the Energy Spectra at SK and SNO

As mentioned in the previous subsection, neutrino oscillation makes $\nu_e$ and $\bar{\nu}_e$ spectra harder. Therefore, the ratio of high-energy events to low-energy events will be a good measure of neutrino oscillation effects. We calculated the following ratio of events at both detectors:

$$R_{SK} \equiv \frac{\text{number of events at } 30 < E < 70\text{MeV}}{\text{number of events at } 5 < E < 20\text{MeV}} \quad (5.1)$$

$$R_{SNO} \equiv \frac{\text{number of events at } 25 < E < 70\text{MeV}}{\text{number of events at } 5 < E < 20\text{MeV}} \quad (5.2)$$

The plots of $R_{SK}$ vs. $R_{SNO}$ are shown in Fig.14. The errorbars include only statistical errors. At first glance,

it seems to be possible to distinguish all the models including the no oscillation case. But there are other ambiguities besides statistical errors.

One is the mass of the progenitor star. Supernovae with different progenitor masses may result in different original neutrino spectra and neutrino oscillation effects. Studies on this point are now in progress. But dependence of shape of neutrino spectra on progenitor mass is not so large [26] and we would be able to distinguish the models. The difference among the following three groups will still be clear: (1)LMA-L and LMA-S, (2)SMA-L, and (3)SMA-S and no oscillation.

Another ambiguity is the direction of supernova. Depending on the direction, neutrinos from supernova may travel through the Earth before they reach the detectors. In this case, we have to take the Earth matter effect into account. This effect has already been studied by our previous work [27]. In this work we concluded that we can differentiate LMA-L from LMA-S, by observing the Earth matter effects.

## VI. SUMMARY

We studied quantitatively the effects of the three-flavor oscillation on supernova neutrinos, by using more realistic neutrino profiles and presupernova density profiles than previous studies. Our calculations are based on a realistic numerical supernova model calculated by the Lawrence Livermore group [17] and a realistic model of density profile of a presupernova star by Woosley and Weaver [20]. First we calculate time evolution of conversion probabilities. Then energy spectra and time evolution of number luminosity are obtained assuming a supernova at 10kpc. Neutronization burst is highly suppressed in models LMA-L and SMA-L. In case that there is neutrino mixing, energy spectra are harder than in case of no oscillation. By comparing ratios of high-energy events to low-energy events at SuperKamiokande and SNO, we found that we will be able to distinguish the solutions of solar neutrino problem and probe $\theta_{13}$.

## ACKNOWLEDGMENTS

We would like to thank S. E. Woosley for giving the numerical data of the progenitor star, and J. R. Wilson and H. E. Dalhed for the neutrino emission data from the supernova explosion. We also want to acknowledge T.Kajita, Y.Totsuka , Y.Suzuki and Y.Fukuda for recent results of neutrino oscillation analysis at SuperKamiokande. This work was supported in part by Grant-in-Aid for Scientific Research provided by the Ministry of Education, Science and Culture of Japan through Research Grant No.07CE2002 and 12047233.

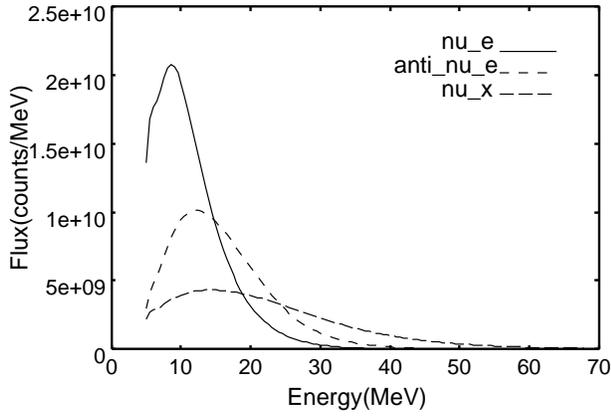

FIG. 1. Original energy spectra of neutrinos [18]. Solid, dashed, and long-dashed lines correspond to $\nu_e$, $\bar{\nu}_e$, and $\nu_x$, respectively.

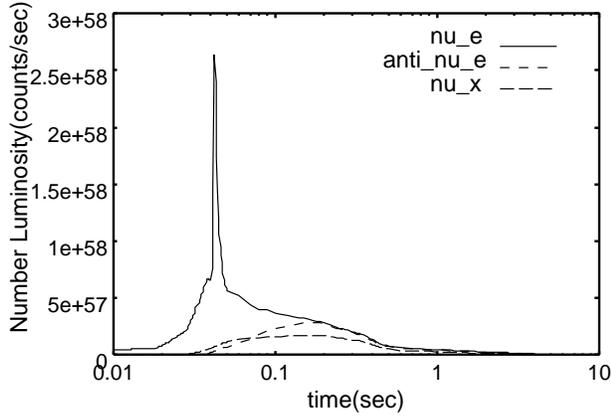

FIG. 2. Time evolution of the original neutrino number luminosity [18]. Solid, dashed, and long-dashed lines correspond to $\nu_e$, $\bar{\nu}_e$, and $\nu_x$, respectively.

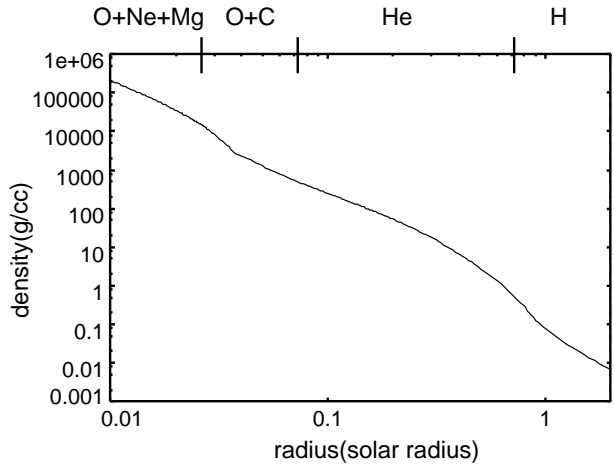

FIG. 3. Density profile of the presupernova star model used in the paper [20]. The progenitor mass is set to be $15 M_\odot$.

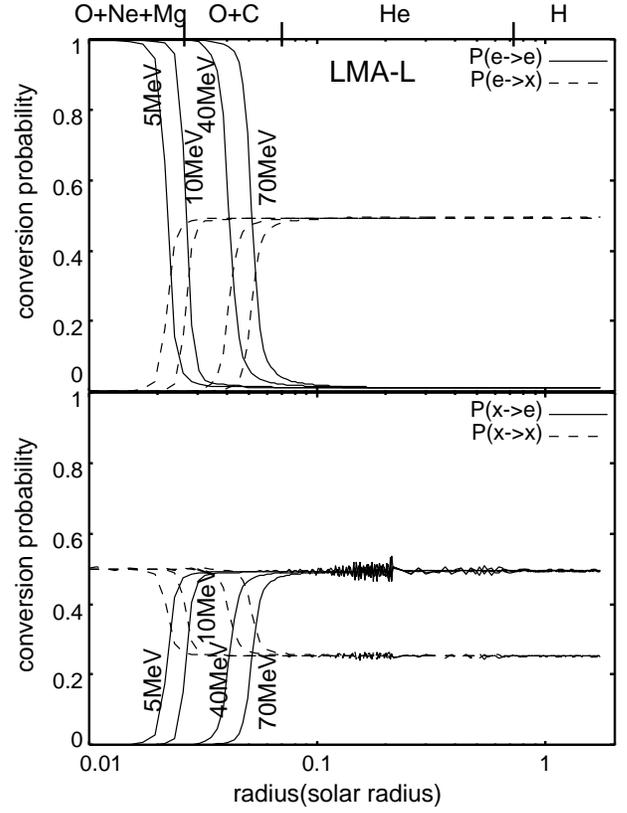

FIG. 4. Time evolution of conversion probability for model LMA-L. In the upper figure, solid and dashed lines show $P(e \rightarrow e)$ and $P(e \rightarrow x)$, respectively. In the lower figure, solid and dashed lines show $P(x \rightarrow e)$ and $P(x \rightarrow x)$, respectively. Four lines of the same marking correspond to neutrino energy, 5MeV, 10MeV, 40MeV, and 70MeV, respectively.

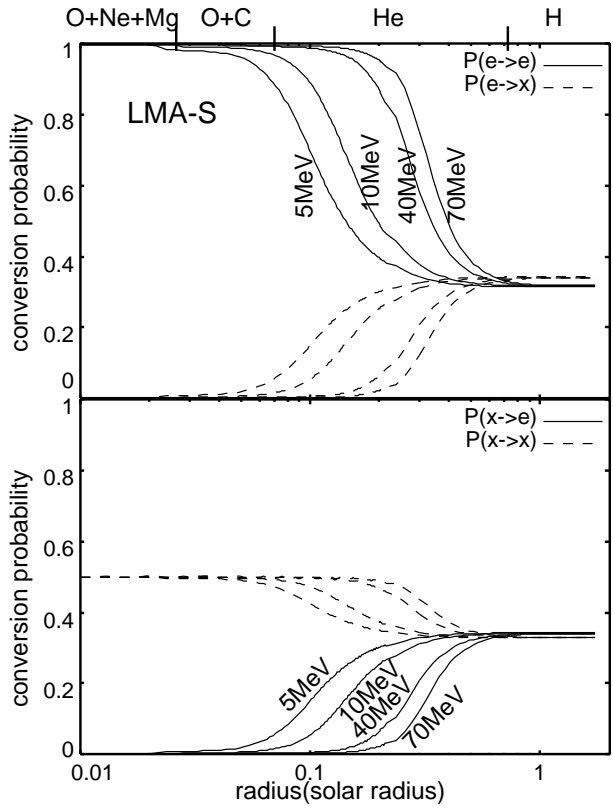

FIG. 5. Time evolution of conversion probability for model LMA-S. In the upper figure, solid and dashed lines show $P(e \to e)$ and $P(e \to x)$, respectively. In the lower figure, solid and dashed lines show $P(x \to e)$ and $P(x \to x)$, respectively. Four lines of the same marking correspond to neutrino energy, 5MeV, 10MeV, 40MeV, and 70MeV, respectively.

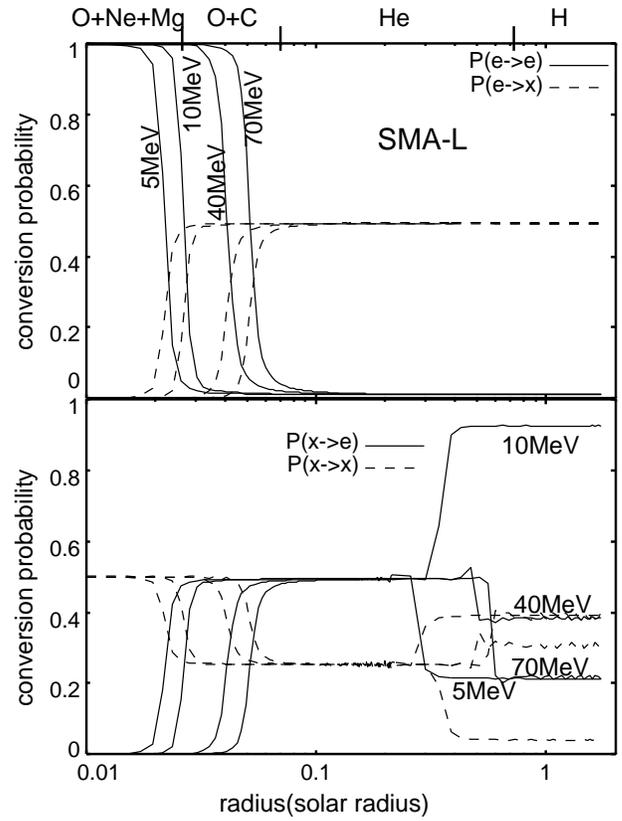

FIG. 6. Time evolution of conversion probability for model SMA-L. In the upper figure, solid and dashed lines show $P(e \to e)$ and $P(e \to x)$, respectively. In the lower figure, solid and dashed lines show $P(x \to e)$ and $P(x \to x)$, respectively. Four lines of the same marking correspond to neutrino energy, 5MeV, 10MeV, 40MeV, and 70MeV, respectively.

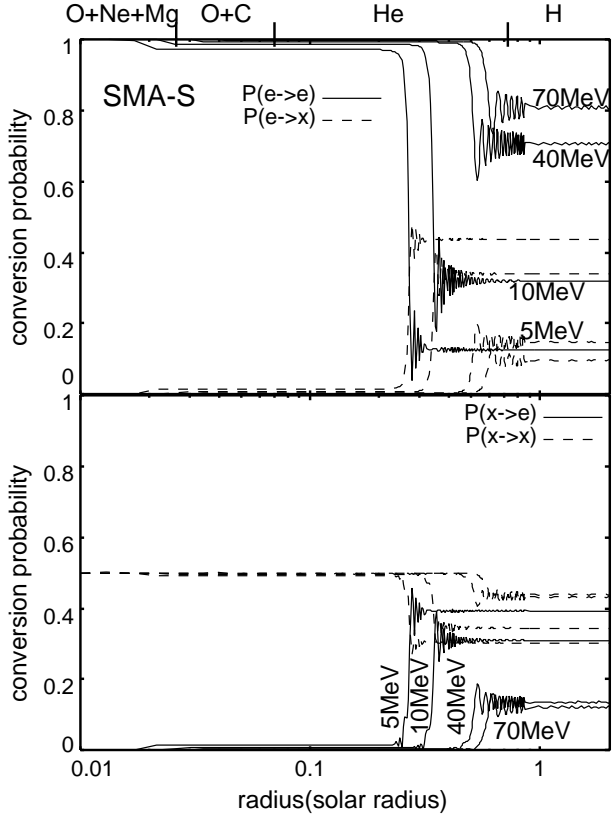

FIG. 7. Time evolution of conversion probability for model SMA-S. In the upper figure, solid and dashed lines show $P(e \to e)$ and $P(e \to x)$, respectively. In the lower figure, solid and dashed lines show $P(x \to e)$ and $P(x \to x)$, respectively. Four lines of the same marking correspond to neutrino energy, 5MeV, 10MeV, 40MeV, and 70MeV, respectively.

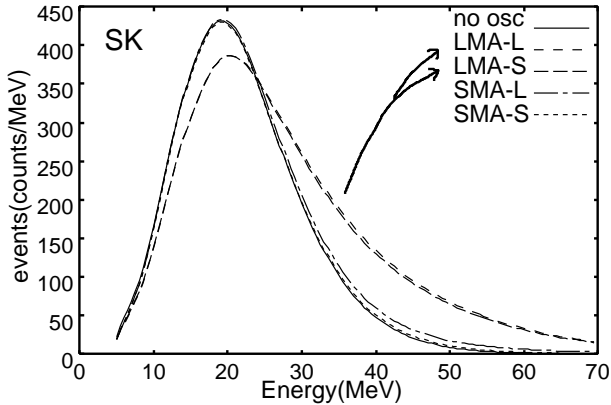

FIG. 8. Energy spectrum of positrons and electrons expected to be detected at SuperKamiokande. Solid, dashed, long-dashed, dash-dot-dash, and dotted lines correspond to no oscillation, model LMA-L, LMA-S, SMA-L, and SMA-S, respectively.

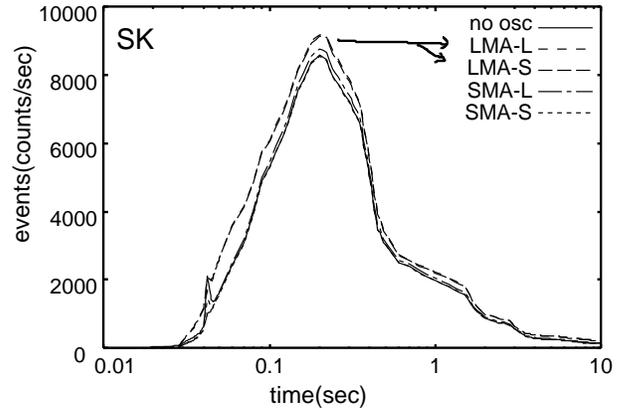

FIG. 9. Time evolution of number luminosity of positrons and electrons expected to be detected at SuperKamiokande. Solid, dashed, long-dashed, dash-dot-dash, and dotted lines correspond to no oscillation, model LMA-L, LMA-S, SMA-L, and SMA-S, respectively.

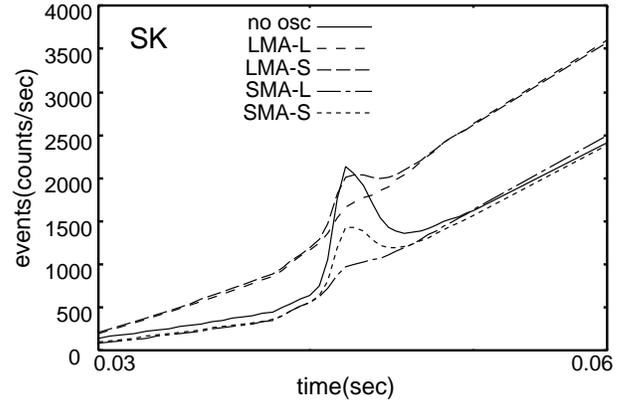

FIG. 10. A zoom-up of Fig.9 near the neutronization burst. Solid, dashed, long-dashed, dash-dot-dash, and dotted lines correspond to no oscillation, model LMA-L, LMA-S, SMA-L, and SMA-S, respectively.

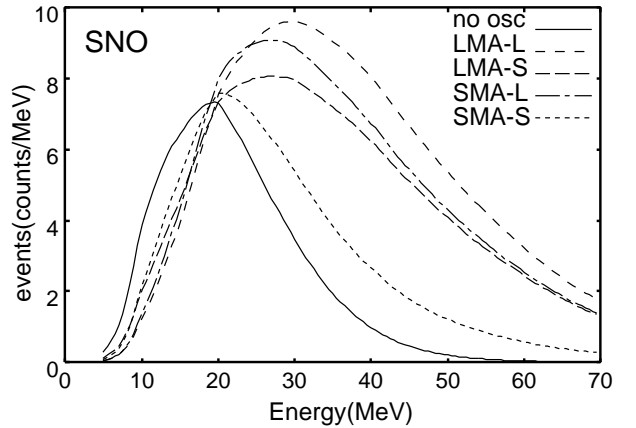

FIG. 11. Energy spectrum of positrons and electrons expected to be detected at SNO taking only CC events into account. Solid, dashed, long-dashed, dash-dot-dash, and dotted lines correspond to no oscillation, model LMA-L, LMA-S, SMA-L, and SMA-S, respectively.

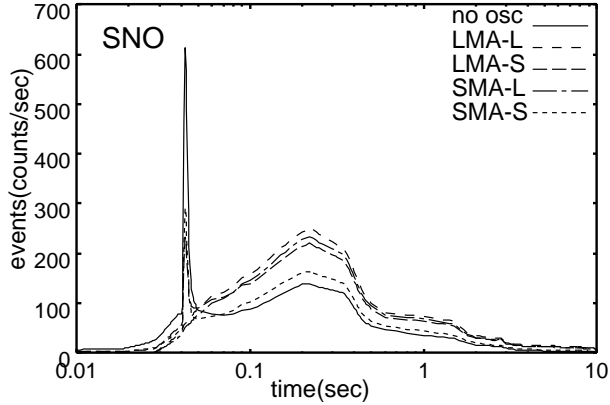

FIG. 12. Time evolution of number luminosity of positrons and electrons expected to be detected at SNO taking only CC events into account. Solid, dashed, long-dashed, dash-dot-dash, and dotted lines correspond to no oscillation, model LMA-L, LMA-S, SMA-L, and SMA-S, respectively.

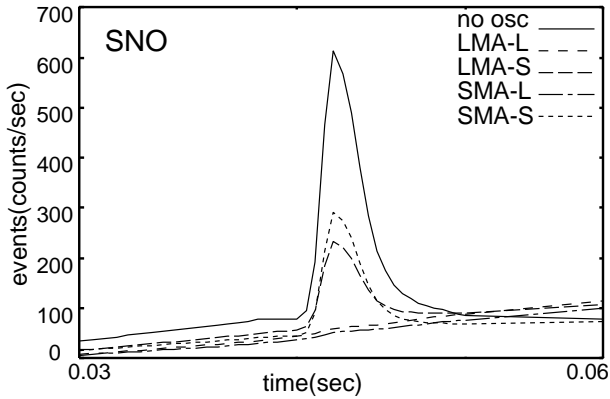

FIG. 13. A zoom-up of Fig.12 near the neutronization burst. Solid, dashed, long-dashed, dash-dot-dash, and dotted lines correspond to no oscillation, model LMA-L, LMA-S, SMA-L, and SMA-S, respectively.

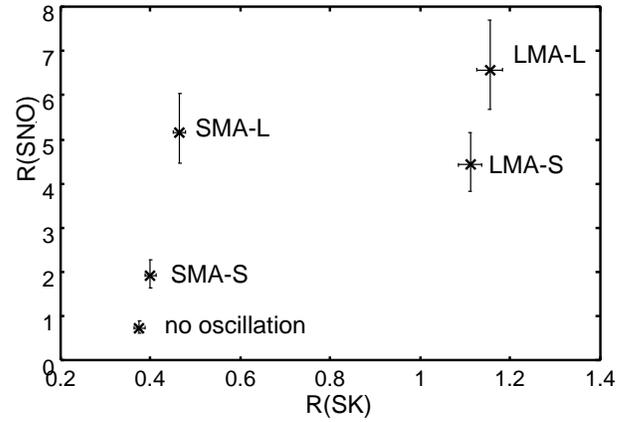

FIG. 14. The plot of $R_{SK}$ vs. $R_{SNO}$ for all the models. The error-bars represent the statistical errors.

TABLE I. Sets of mixing parameter for calculation

| model | $\sin^2 2\theta_{12}$ | $\sin^2 2\theta_{23}$ | $\sin^2 2\theta_{13}$ | $\Delta m_{12}^2 (\text{eV}^2)$ | $\Delta m_{13}^2 (\text{eV}^2)$ | $\nu_\odot$ problem | H resonance | L resonance |
|---|---|---|---|---|---|---|---|---|
| LMA-L | 0.87 | 1.0 | 0.043 | $7.0 \times 10^{-5}$ | $3.2 \times 10^{-3}$ | LMA | adiabatic | adiabatic |
| LMA-S | 0.87 | 1.0 | $1.0 \times 10^{-6}$ | $7.0 \times 10^{-5}$ | $3.2 \times 10^{-3}$ | LMA | nonadiabatic | adiabatic |
| SMA-L | $5.0 \times 10^{-3}$ | 1.0 | 0.043 | $6.0 \times 10^{-6}$ | $3.2 \times 10^{-3}$ | SMA | adiabatic | nonadiabatic |
| SMA-S | $5.0 \times 10^{-3}$ | 1.0 | $1.0 \times 10^{-6}$ | $6.0 \times 10^{-6}$ | $3.2 \times 10^{-3}$ | SMA | nonadiabatic | nonadiabatic |

TABLE II. Number of events at SuperKamiokande

| model | LMA-L | LMA-S | SMA-L | SMA-S | no osc |
|---|---|---|---|---|---|
| $\bar{\nu}_e p$ | 9459 | 9427 | 8101 | 7967 | 8036 |
| $\nu_e e^-$ | 186 | 115 | 189 | 131 | 132 |
| $\bar{\nu}_e e^-$ | 46 | 46 | 41 | 42 | 42 |
| $\nu_\mu e^-$ | 25 | 26 | 25 | 30 | 30 |
| $\bar{\nu}_\mu e^-$ | 24 | 23 | 24 | 24 | 24 |
| $\nu_\tau e^-$ | 25 | 26 | 25 | 30 | 30 |
| $\bar{\nu}_\tau e^-$ | 24 | 23 | 24 | 24 | 24 |
| $O\nu_e$ | 297 | 214 | 297 | 108 | 31 |
| $O\bar{\nu}_e$ | 160 | 158 | 95 | 92 | 92 |
| total | 10245 | 10114 | 8822 | 8447 | 8441 |
| neutronization burst | 15.7 | 16.7 | 9.0 | 10.1 | 12.4 |

TABLE III. Number of events (CC) at SNO

| model | LMA-L | LMA-S | SMA-L | SMA-S | no osc |
|---|---|---|---|---|---|
| $\nu_e d$(CC) | 237 | 185 | 237 | 111 | 68 |
| $\bar{\nu}_e d$(CC) | 118 | 117 | 84 | 82 | 82 |
| total | 355 | 302 | 321 | 193 | 150 |
| neutronization burst | 0.6 | 1.1 | 0.5 | 1.1 | 2.1 |